\documentclass[12pt]{article}

\usepackage{natbib,graphicx}

\begin{document}

\title{The Kepler map in the three-body problem}

\author{Ivan I. Shevchenko \\
Pulkovo Observatory of the Russian Academy of Sciences, \\
Pulkovskoje ave. 65-1, St.\,Petersburg 196140, Russia \\
Email: iis@gao.spb.ru}

\maketitle

\begin{abstract}
The Kepler map was derived by Petrosky (1986) and Chirikov and
Vecheslavov (1986) as a tool for description of the long-term
chaotic orbital behaviour of the comets in nearly parabolic
motion. It is a two-dimensional area-preserving map, describing
the motion of a  comet in terms of energy and time. Its second
equation is based on Kepler's third law, hence the title of the
map. Since 1980s the Kepler map has become paradigmatic in a
number of applications in celestial mechanics and atomic physics.
It represents an important kind of general separatrix maps.
Petrosky and Broucke (1988) used refined methods of mathematical
physics to derive analytical expressions for its single parameter.
These methods became available only in the second half of the 20th
century, and it may seem that the map is inherently a very modern
mathematical tool. With the help of the Jacobi integral I show
that the Kepler map, including analytical formulae for its
parameter, can be derived by quite elementary methods. The
prehistory and applications of the Kepler map are considered and
discussed. \linebreak Keywords: celestial mechanics, stellar
dynamics, methods: analytical, methods: n-body simulations.
\end{abstract}

\section{Introduction}
\label{intro}

The Kepler map was discovered in 1986 by physicists, but in
application to dynamical astronomy. The first publications
belonged to \cite{P86} and \cite{CV86}; they were followed very
soon by many other contributions, where, on one hand, the initial
results were further developed and described in much greater
detail \citep{PB88,VC88,CV89,E90} and, on the other hand, the same
mathematical construction was derived in application to problems
in atomic physics \citep{CGS87,GK87,CGS88,BGS88,JLR88}.

\cite{P86} and \cite{CV86} derived the Kepler map as a tool for
description of the chaotic motion of the comets in near-parabolic
orbits. The model consists in the assumption that the main
perturbing effect of a planet is concentrated when the comet is
close to the perihelion of its orbit. This effect is defined by
the phase of encounter with the planet.

Today the Kepler map is known and used to describe dynamics in
several different settings of a hierarchical three-body problem:
in the external restricted planar \citep{P86,PB88} and strongly
non-planar \citep{E90} problems in cometary dynamics; as
well as in the abstract Sitnikov problem, where the tertiary moves
{\it perpendicular} to the orbital plane of the main binary.
(\cite{UH08} considered a variant of the Sitnikov problem and
derived a map, which is in fact the Kepler map; see Eqs.~(11) in
their paper.)

The Kepler map has a single parameter. Its analytical formula was
first given (in the restricted planar three-body problem) by
\cite{P86}, but only in a simplified form of asymptotics for large
values of the pericentre distance of the cometary orbit, and the
deduction was not attached. The latter was provided by
\cite{PB88}. They used refined methods of mathematical physics to
derive analytical expressions for the parameter. These methods
became available only in the second half of the 20th century, and
it may seem that the map is inherently a very modern mathematical
tool. However, in the present paper I show that the Kepler map,
including analytical formulae for its parameter, can be derived by
quite elementary methods. What is more, the asymptotics for its
parameter can be obtained by a method, which is much simpler than
that used in \citep{PB88}. I discuss the prehistory of the Kepler
map and its current applications, and demonstrate that the
necessary tools for the derivation of the Kepler map have become
available already in the middle of 19th century.

\section{Elementary derivation of the Kepler map}
\label{edkm}

Let us consider the motion of a comet in the planar restricted
three-body problem Sun--Jupiter--comet. We choose an inertial
Cartesian coordinate system with the origin at the mass centre of
the Sun and Jupiter. The motion of a comet with the coordinates
$(x, y)$ is described by the differential equations
\begin{eqnarray}
\ddot x &=& \nu \frac{x_{\rm S} - x}{r^3_{13}} +
\mu \frac{x_{\rm J} - x}{r^3_{23}} , \nonumber \\
\ddot y &=& \nu \frac{y_{\rm S} - y}{r^3_{13}} + \mu \frac{y_{\rm
J} - y}{r^3_{23}}
\label{diffeqs}
\end{eqnarray}

\noindent (see, e.g., \cite{S67}), where
\begin{eqnarray}
r^2_{13} &=& (x_{\rm S} - x)^2 + (y_{\rm S} - y)^2 , \nonumber \\
r^2_{23} &=& (x_{\rm J} - x)^2 + (y_{\rm J} - y)^2 ,
\label{rr}
\end{eqnarray}
\begin{eqnarray}
x_{\rm S} &=& - \mu \cos(t - t_0) , \nonumber \\
y_{\rm S} &=& - \mu \sin(t - t_0) ,
\label{xyS}
\end{eqnarray}
\begin{eqnarray}
x_{\rm J} &=& \nu \cos(t - t_0) , \nonumber \\
y_{\rm J} &=& \nu \sin(t - t_0) ,
\label{xyJ}
\end{eqnarray}

\noindent where $r_{13}$ and $r_{23}$ are the distances Sun--comet
and Jupiter--comet, respectively; $(x_{\rm S}, y_{\rm S})$ and
$(x_{\rm J}, y_{\rm J})$ are the coordinates of the Sun and
Jupiter, respectively; $\mu$ is the mass of Jupiter, $\nu = 1 -
\mu$ is the mass of the Sun. We set the length unit to be equal to
the constant Sun--Jupiter distance, the mass unit equal to the sum
of the Solar and Jovian masses, and the time unit equal to $1/(2
\pi)$ of the period of the orbital motion of Jupiter.

Let us expand the right-hand sides of Eqs.~(\ref{diffeqs}) in
power series of $\mu$, retaining the first-order terms only:
\begin{eqnarray}
\ddot x &=& - \frac{x}{r^3} + \mu F(x, y, t, t_0), \nonumber \\
\ddot y &=& - \frac{y}{r^3} + \mu G(x, y, t, t_0),
\label{deappr}
\end{eqnarray}

\noindent with
$r = (x^2 + y^2)^{1/2}$,
\begin{eqnarray}
F(x, y, t, t_0) &=& [ x - \cos(t - t_0) ] r^{-3} +
3x [ x \cos(t - t_0) + y \sin(t - t_0) ] r^{-5} + \nonumber \\
&+& [ \cos(t - t_0) - x ] \{[x - \cos(t - t_0)]^2 + [y - \sin(t - t_0)]^2\}^{-3/2} ,
\qquad \label{Ftt0}
\end{eqnarray}
\begin{eqnarray}
G(x, y, t, t_0) &=& [ y - \sin(t - t_0) ] r^{-3} +
3y[x \cos(t - t_0) + y \sin(t - t_0)] r^{-5} + \nonumber \\
&+& [ \sin(t - t_0) - y ] \{[x - \cos(t - t_0)]^2 + [y - \sin(t - t_0)]^2\}^{-3/2} ;
\qquad \label{Gtt0}
\end{eqnarray}

\noindent see, e.g., \citep{LS94,Z00}. The quantity $t_0$ is the
initial epoch. It is chosen in such a way, that the comet is at
the perihelion of its orbit when $t = 0$. Designating the phase
angle of Jupiter at $t = 0$ as $g$, one has $t_0 = - g$.

The constant energy $E$ of the unperturbed orbital motion is
\begin{equation}
E = \frac{1}{2}( \dot x^2 + \dot y^2 ) - \frac{1}{r} = -
\frac{1}{2 a} ,
\label{Kmu0}
\end{equation}

\noindent where $a$ is the semi-major axis of the cometary orbit.
For the perturbed motion the energy is
\begin{equation}
E = \frac{1}{2}( \dot x^2 + \dot y^2 ) - \frac{1-\mu}{r_{13}} -
\frac{\mu}{r_{23}} = - \frac{1}{2 a}
\label{Kmu}
\end{equation}

\noindent \citep{S67}, and it is not constant. Then from
Eqs.~(\ref{deappr}) one has
\begin{equation}
\dot E = \mu [ \dot x(t) F(x(t), y(t), t, t_0) + \dot y(t) G(x(t), y(t), t, t_0)].
\end{equation}

\noindent The increment of the energy $E$ for one cometary orbital
period is given by the integral \citep{LS94,Z00}:
\begin{equation}
\Delta E = \mu \int_{-\infty}^{+\infty} [ \dot x(t)F(t, g)+ \dot y(t)G(t, g)] {\rm d}t ,
\end{equation}

\noindent where $g = -t_0$. $\Delta E$ is a $2 \pi$-periodic
function of $g$. It is anti-symmetric with respect to $g=\pi$.

In the inertial coordinate system that we have chosen (that with
the origin at the mass centre of the Sun and Jupiter), the Jacobi
integral is
\begin{equation}
\dot x^2 + \dot y^2 - \frac{2 (1 - \mu)}{r_{13}} - \frac{2
\mu}{r_{23}} - 2 ( x \dot y - y \dot x ) = \mbox{const}
\label{JI}
\end{equation}

\noindent \citep{S67}, or, $E - D = \mbox{const}$. So, $\dot E =
\dot D$. Let us derive an analytical expression for the increment
of the angular momentum $D$ per one orbital revolution of the
tertiary. We choose the angular momentum, because in the case of
the energy the analytical calculation is too complicated to
achieve the result; however, as we have just seen, the result must
be the same. The angular momentum is
\begin{equation}
D =  x \dot y - y \dot x ,
\label{D}
\end{equation}

\noindent and its time derivative
\begin{equation}
\dot D =  x \ddot y - y \ddot x .
\label{dD}
\end{equation}

\noindent Substituting Eqs.~(\ref{diffeqs}) for $\ddot y$ and
$\ddot x$, one has

\begin{equation}
\dot D = \nu \frac{x y_{\rm S} - x_{\rm S} y}{r^3_{13}} +
\mu \frac{x y_{\rm J} - x_{\rm J} y}{r^3_{23}} ,
\label{evaldD}
\end{equation}

\noindent where
\begin{eqnarray}
r^2_{13} &=& \mu^2 + r^2 - 2 (x_{\rm S} x + y_{\rm S} y) , \nonumber \\
r^2_{23} &=& \nu^2 + r^2 - 2 (x_{\rm J} x + y_{\rm J} y) ,
\label{rrm}
\end{eqnarray}

\noindent i.e., $\dot D$ is the sum of four terms:
\begin{equation}
\dot D =  {\cal A} + {\cal B} + {\cal C} + {\cal D}
\label{dD4}
\end{equation}

\noindent with
\begin{equation}
{\cal A} = \nu \frac{x y_{\rm S}}{r^3_{13}} , \
{\cal B} = - \nu \frac{x_{\rm S} y}{r^3_{13}} , \
{\cal C} = \mu \frac{x y_{\rm J}}{r^3_{23}} , \
{\cal D} = - \mu \frac{x_{\rm J} y}{r^3_{23}} .
\label{abcd}
\end{equation}

\noindent It is sufficient to find ${\cal A}$ and ${\cal B}$, because
\begin{equation}
{\cal C} = - {\cal A} (\nu \to - \mu) , \ {\cal D} = - {\cal B} (\nu \to - \mu) .
\label{abcd_id}
\end{equation}

\noindent Let us write down the well-known elementary formulae for
the unperturbed parabolic motion:
\begin{equation}
r = q (1 + u^2) , \
x = q \left(1 - u \right), \
y = 2 q u, \
t = \kappa \left( u + \frac{u^3}{3} \right), \
\label{parabolic}
\end{equation}

\begin{equation}
u = \left(\tau + (1+\tau^2)^{1/2} \right)^{1/3} +
\left(\tau - (1+\tau^2)^{1/2} \right)^{1/3}, \
\tau = \frac{3}{2 \kappa} t ,
\label{parabolic1}
\end{equation}

\noindent where $\kappa = (2 q^3)^{1/2}$, the eccentric anomaly $u
= \tan \frac{f}{2}$, and $q$ and $f$ are the perihelion distance
and the true anomaly, respectively. Note that we consider solely
the case of prograde orbits here; analysis of the retrograde case
is analogous.

We use exact relations~(\ref{parabolic}) for substitutions in
calculating the increment of the angular momentum. Thus we follow
a standard approach for deriving the energy increments in the
separatrix maps \citep{C79}. So, inserting Eqs.~(\ref{parabolic})
in Eqs.~(\ref{abcd}), we find
\begin{eqnarray}
{\cal A} &=& - \mu \nu q \frac{(1 - u^2)
\sin \left[ \kappa \left( u + \frac{u^3}{3} \right) - t_0 \right]}
{r^3_{13}} , \nonumber \\
{\cal B} &=& 2 \mu \nu q \frac{u
\cos \left[ \kappa \left( u + \frac{u^3}{3} \right) - t_0 \right]}
{r^3_{13}} .
\label{ab}
\end{eqnarray}

\noindent Combining Eqs.~(\ref{rrm}), (\ref{xyS}), and (\ref{xyJ})
and inserting Eqs.~(\ref{parabolic}) in the resulting expressions,
we find an expression for $r_{13}$ to substitute in the
denominators in Eqs.~(\ref{ab}):
\begin{eqnarray}
r^2_{13} &=& \mu^2 + q^2 (1 + u^2)^2 + 2 \mu q \left\{ (1 - u^2)
\cos \left[ \kappa \left( u + \frac{u^3}{3} \right) - t_0 \right]
+ 2 u \sin \left[ \kappa \left( u + \frac{u^3}{3} \right) - t_0
\right] \right\} . \nonumber \\
&{}&
\label{rrmc}
\end{eqnarray}

\noindent Then, expanding the right-hand sides of Eqs.~(\ref{ab})
in power series in $\mu$, taking into account that $q \gg 1$, we
obtain in the first order of $\mu$:
\begin{eqnarray}
{\cal A} + {\cal C} & = & - \frac{3 \mu}{2 q^4}
\frac{(1 - u^2) \sin \left[ \kappa \left( u + \frac{u^3}{3} \right) - t_0 \right]}
{(1 + u^2)^5} , \nonumber \\
{\cal B} + {\cal D} & = & \frac{3 \mu}{q^4}
\frac{u \cos \left[ \kappa \left( u + \frac{u^3}{3} \right) - t_0 \right]}
{(1 + u^2)^5} .
\label{abmu}
\end{eqnarray}

From Eqs.~(\ref{dD4}) and (\ref{parabolic}) one can find the
angular momentum increment per an orbital revolution of the comet.
As follows from the Jacobi integral~(\ref{JI}), the angular
momentum increment is equal to the energy increment. So, the
energy increment is given by the integral
\begin{equation}
\Delta E =
\kappa \int_{-\infty}^{+\infty} ({\cal A} + {\cal B} + {\cal C} + {\cal D})
(1 + u^2) {\rm d}u .
\label{deltaE1}
\end{equation}

\noindent To evaluate it, first of all we define the functions
\begin{eqnarray}
I_n^0 (x) & = & \int_{-\infty}^{+\infty} \frac{1}{(1 + u^2)^n}
\cos \left[ x \left( u + \frac{u^3}{3} \right) \right] {\rm d}u , \nonumber \\
I_n^1 (x) & = & \int_{-\infty}^{+\infty} \frac{u}{(1 + u^2)^n}
\sin \left[ x \left( u + \frac{u^3}{3} \right) \right] {\rm d}u , \nonumber \\
I_n^2 (x) & = & \int_{-\infty}^{+\infty} \frac{u^2}{(1 + u^2)^n}
\cos \left[ x \left( u + \frac{u^3}{3} \right) \right] {\rm d}u . \nonumber \\
\label{Idef}
\end{eqnarray}

\noindent Two of them, $I_n^0$ and $I_n^1$, were introduced by
\cite{PB88} in a different designation. The following recurrent
relations
\begin{eqnarray}
I_{n+1}^1 (x) & = & \frac{x}{2 n} I_{n-1}^0 (x) , \nonumber \\
2 n I_{n+1}^0 (x) & = & (2 n - 1) I_n^0 (x) + x I_{n-1}^1 (x), \nonumber \\
I_n^2 (x) & = & I_{n-1}^0 (x) - I_n^0 (x), \nonumber \\
\frac{\mathrm{d} I_n^0 (x)}{\mathrm{d} x} & = & - \frac{2}{3} I_n^1 (x) - \frac{1}{3} I_{n-1}^1 (x) , \nonumber \\
\frac{\mathrm{d} I_n^1 (x)}{\mathrm{d} x} & = & - \frac{2}{3} I_n^0 (x) + \frac{1}{3} I_{n-1}^0 (x) + \frac{1}{3} I_{n-2}^0 (x)
\label{Irecrel}
\end{eqnarray}

\noindent are valid for these functions (the 1st, 2nd, 4th, and
5th of them were deduced and used by \cite{PB88} in other
designations; see appendix in their paper).

From Eq.~(\ref{deltaE1}) we find
\begin{equation}
\Delta E = W(q) \sin t_0 ,
\label{deltaE}
\end{equation}

\noindent where
\begin{eqnarray}
W(q) & = & \frac{3 \mu}{2^{1/2} q^{5/2}}
[ I_4^0 (\kappa) + 2 I_4^1 (\kappa) - I_4^2 (\kappa) ] = \nonumber \\
         & = & \frac{3 \mu}{2^{1/2} q^{5/2}}
[ 2 I_4^0 (\kappa) + 2 I_4^1 (\kappa) - I_3^0 (\kappa) ] ,
\label{Wq}
\end{eqnarray}

\noindent where $\kappa = (2 q^3)^{1/2}$. This coefficient, if
divided by 4, coincides with the corresponding coefficient found
by \cite{PB88}; see the last equation in the appendix in their
paper. Most probably, the deviation of factor 4 is due to a
misprint in their paper, because the final asymptotic results,
compared below, coincide completely.

As demonstrated by \cite{PB88}, some of the terms in
Eq.~(\ref{Wq}) can be expressed through the modified Bessel
functions of the second kind and the Airy functions, because
\begin{equation}
I_0^0 (x) = 3^{-1/2} K_{1/3} \left( \frac{2}{3}x \right) =
\pi x^{-1/3} {\rm Ai} \left( x^{2/3} \right) , \ \
I_0^1 (x) = 3^{-1/2} K_{2/3} \left( \frac{2}{3}x \right) ,
\label{IvBf}
\end{equation}

\noindent where
\begin{eqnarray}
K_\nu (x) & = & \sec \left( \frac{1}{2} \nu \pi \right)
\int_0^\infty
\cos (x \sinh t) \cosh (\nu t) {\rm d}t , \nonumber \\
{\rm Ai} (x) & = & \frac{1}{\pi} \int_0^{\infty}
\cos \left( x t + \frac{t^3}{3} \right) {\rm d}t
\label{Bfdef}
\end{eqnarray}

\noindent by definition, see \citep{AS70,PB88}. However, the
$W(q)$ coefficient has not yet been completely expressed through
known special functions. \cite{P86} and \cite{PB88} found a
formula for the asymptotics of $W(q)$ at $q \to \infty$. Its
derivation is given in the appendix of \citep{PB88}. It is rather
complicated and involves, in particular, approximate analytical
solution of an ancillary differential equation and approximate
numerical evaluation of an integral.

Here we show that the asymptotics can be derived in a much more
straightforward and simple way. First of all, using the 1st, 2nd,
3rd, and 4th recurrent relations in list~(\ref{Irecrel}), we
reduce Eq.~(\ref{Wq}) to the form
\begin{equation}
W(q) = \frac{3 \mu}{2^{1/2} q^{5/2} \kappa}
\left[ 2 I_6^0 (\kappa) + 36 I_6^1 (\kappa) - 18 I_6^2 (\kappa) +
24 \frac{\mathrm{d} I_6^0 (\kappa)}{\mathrm{d} \kappa} \right] .
\label{Wq6}
\end{equation}

\noindent Then we take the asymptotic expressions for $I_6^n (x)$
($n=0$, 1, 2) at $x \to \infty$ from the papers \citep{H75,RH03},
where the corresponding integral was evaluated using the method of
steepest descents (the only complication in using this method was
that the saddle points of the exponent under integral are situated
at the poles of the integrand). These expressions are
\begin{equation}
I_6^0 (x) \simeq I_6^1 (x) \simeq - I_6^2 (x) \simeq
\frac{\pi^{1/2}}{120} x^{5/2} \exp \left( - \frac{2}{3} x \right) .
\label{IRH}
\end{equation}

\noindent Finally, we arrive at
\begin{equation}
W(q) \simeq 2^{1/4} \pi^{1/2} \mu q^{-1/4} \exp \left( - \frac{(2
q)^{3/2}}{3} \right) ,
\label{Wqasy}
\end{equation}

\noindent in complete agreement with formula (3.16{\it a}) in
\citep{PB88}. (Except that the minus sign is obviously lacking
under the exponent in eq.~(3.16{\it a}) in \citep{PB88}, due to a
misprint. Note that the same coefficient given in \citep{P86} is
$2 \pi$ times greater; this is apparently a misprint.)

\section{The Kepler map: limits for application}
\label{kmla}

As \cite{P86} discovered, if one writes down the expression for
the energy increment together with the expression for the
increment of Jupiter's phase angle $g$ (following from Kepler's
third law) on the time interval ``between two consecutive
perihelion passages'' (as it is usually stated), one obtains a
two-dimensional area-preserving map
\begin{eqnarray}
E_{i+1} &=& E_{i} + W(q) \sin g_{i} , \nonumber \\
g_{i+1} &=& g_{i} + 2 \pi \vert 2 E_{i+1} \vert^{-3/2} ,
\label{gm2}
\end{eqnarray}

\noindent where the subscript $i$ denotes the current number of
the perihelion passage, $g_i = -t_0$. The coefficient $W(q)$ is
given by formulae~(\ref{Wq6}) and (\ref{Wqasy}), if $\mu \ll 1$
and $q \gg 1$.

In fact, there is an inconsistency here: instead of formulation
``between two consecutive perihelion passages'', it is correct to
say that the energy increment in Eqs.~(\ref{gm2}) is taken between
two consecutive {\it aphelion} passages, while the phase increment
is indeed taken between perihelion passages. In atomic physics,
this inconsistency was pointed out by \cite{N90}, who derived a
more complicated map without this asynchronism. The asynchronism
can be as well removed without construction of a separate map, but
by means of a simple procedure of synchronization, described for
the case of ordinary separatrix maps in \citep{S98b,S00}.

By means of substitution $E = W y$, $g = x$, map~(\ref{gm2}) is
reducible to
\begin{eqnarray}
     y_{i+1} &=& y_i + \sin x_i, \nonumber \\
     x_{i+1} &=& x_i + \lambda \vert y_{i+1} \vert^{-3/2} ,
\label{gm1}
\end{eqnarray}

\noindent  where $\lambda = 2^{-1/2} \pi W^{-3/2}$. The $y$
variable has the meaning of the normalized orbital energy of the
comet, and $x$ is the normalized time.

Since $W \ll 1$ usually (see Eq.~(\ref{Wqasy})), one has $\lambda
\gg 1$. This means that chaos in the motion of comets is not
adiabatic. In particular, the Kepler map can be locally
approximated by the standard map with good accuracy.

One iteration of the Kepler map corresponds to one orbital
revolution of the comet, and this means that the map time unit,
corresponding to one iteration, is not constant. The increment of
real time per iteration is $\Delta x_{i+1} = x_{i+1} - x_i$.

In the considered model, the pericentre distance $q$ is set to be
constant. This was justified in \citep{LS94}: they showed that the
variation of $q$, at each return of a comet, if $q \gg 1$, affects
the value of $\Delta E$ only in the second order of $\mu$.
According to \citep{PB88}, the higher order harmonics in $\Delta
E$ are exponentially small with $q$ with respect to the first
harmonic.

If $q > 1$, as in the case considered above, then the comet does
not cross the orbit of Jupiter. If $q < 1$, the orbit of Jupiter
is crossed and $\Delta E$ as a function of $g$ has two
singularities with $| \Delta E | \to \infty$; see
\citep{Z00,ZSZ02}.

\section{The Kepler map as a general separatrix map}
\label{kmsm}

The Kepler map is an example of a general separatrix map. In its
model, the separatrix (the $y=0$ line) separates the bound and
unbound states of motion.

As distinct from the Kepler map, the well-known ordinary
separatrix map has a logarithmic, with respect to the energy,
increment in phase. To ensure a direct comparison with the Kepler
map~(\ref{gm1}), let us write down the ordinary separatrix map in
the form adopted in \citep{S98a}:
\begin{eqnarray}
     y_{i+1} &=& y_i + \sin x_i, \nonumber \\
     x_{i+1} &=& x_i + \lambda \ln \vert y_{i+1} \vert + c ,
\label{sm}
\end{eqnarray}

\noindent where $\lambda$ and $c$ are parameters. In the perturbed
pendulum model of nonlinear resonance, $y$ denotes the normalized
relative pendulum's energy, $x$ is normalized time.

Consider a map similar to map~(\ref{sm}), but with a power-law
phase increment instead of the logarithmic one:
\begin{eqnarray}
     y_{i+1} &=& y_i + \sin x_i, \nonumber \\
     x_{i+1} &=& x_i + \lambda \vert y_{i+1} \vert^{-\gamma} ,
\label{km}
\end{eqnarray}

\noindent or, in an equivalent form commonly used,
\begin{eqnarray}
     w_{i+1} & = & w_i + W \sin \tau_i , \nonumber \\
     \tau_{i+1} & = & \tau_i + \nu \vert w_{i+1} \vert^{-\gamma} .
     \label{km1}
\end{eqnarray}

\noindent Map (\ref{km1}) has two parameters, $W$ and $\nu$,
instead of the single parameter $\lambda$ in map (\ref{km}); apart
from the $\gamma$ parameter. The two-parameter map~(\ref{km1}) is
reducible to the one-parameter map~(\ref{km}) with $\lambda = \nu
| W |^{-\gamma}$ by means of the substitution $w = W y$, $\tau =
x$.

A number of mechanical and physical models are described by maps
(\ref{km}) and (\ref{km1}) with rational values of $\gamma$. The
values of $\gamma = 1/4$ and $1/3$ correspond to the Markeev maps
\citep{M95,M94} for the motion near the separatrices of resonances
in two degenerate cases; $\gamma = 1/2$ gives the ``$\hat L$-map''
\citep{ZSUC91} for the motion of a non-relativistic particle in
the field of a wave packet; this value of $\gamma$ also gives a
map for the classical Morse oscillator driven by time-periodic
force \citep{A06}; $\gamma = 1$ gives the Fermi map
\citep{ZC64,LL92} for the Fermi acceleration mechanism for cosmic
rays; $\gamma = 3/2$ gives the Kepler map for a number of
astronomical and physical applications; $\gamma = 2$ gives the
``ultrarelativistic map'' \citep{ZSUC91} for the motion of a
relativistic particle in the field of a wave packet.

\section{Applications of the Kepler map in dynamical astronomy}

Major modern domains of application of the Kepler map in dynamical
astronomy are as follows.

\begin{itemize}

\item Highly eccentric motion in the restricted planar three-body
problem without crossings of orbits of planets \citep{P86,PB88}.

\item Highly eccentric motion in the restricted non-planar
three-body and four-body problems with crossings of orbits of
planets \citep{CV86,VC88,CV89}.

\item Mean-motion resonances in the perturbed highly eccentric
motion \citep{P86,MT99,PS04}.

\item Chaotic diffusion in the dynamics of comets and meteor
streams \citep{E90,E92,LS94,ZS01,Z00,ZSZ02,MT99}.

\item The Sitnikov problem \citep{UH08}.

\end{itemize}

The Kepler map was invented as a tool for exploring the chaotic
dynamics of particles in the perturbed highly elongated orbits.
This is already clear from the titles of the pioneering works:

\noindent \cite{P86}: ``Chaos and cometary clouds in the Solar
system'';

\noindent \cite{CV86}: ``Chaotic dynamics of comet Halley'';

\noindent \cite{SZ87}: ``Stochasticity in the Kepler problem and a
model of possible dynamics of comets in the Oort cloud'';

\noindent \cite{PB88}: ``Area-preserving mappings and
deterministic chaos for nearly parabolic motion'';

\noindent \cite{CV89}: ``Chaotic dynamics of comet Halley''.

Since the discovery of the Kepler map by \cite{P86} and
\cite{CV86}, the most important generalization of the Kepler map
(already performed heuristically in a first approximation by
\cite{CV86}) has been an introduction of a ``non-harmonic'' Kepler
map, where the energy increment is a truncated series of Fourier
harmonics in the phase variable, or it is a tabulated periodic
function, which may have singularities. This allows one to
describe the cometary motion with $q$ close to one and even less
than one. In the planar circular restricted three-body problem
Sun--planet--comet, \cite{LS94} derived a non-harmonic Kepler
map describing the dynamical evolution of comets in near-parabolic
orbits under the perturbation of a planet, when $q$ can be close
to 1. \cite{Z00} generalized this approach for the planet-crossing
case, when $q$ can be less than 1.

\section{Applications of the Kepler map in physics}

Major modern domains of application of the Kepler map in physics
are as follows.

\begin{itemize}

\item Classical chaotic ionization of hydrogen atoms in a
microwave field \citep{GK87,CGS88,JLR88,JSS91}.

\item Generalizations of the Kepler map for multi-frequency fields
\citep{KV99}.

\item Hydrogen atoms driven by microwave with arbitrary
polarization \citep{PZ01}.

\item The ``synchronized'' Kepler map \citep{N90,PZ01}.

\end{itemize}

\medskip

Similar to the astronomical applications, the Kepler map was
invented as a tool for exploring a chaotic behaviour, as it is
clear from the contents of the pioneering works, which appeared
practically in the same time as in astronomy:

\noindent \cite{GK87}: ``Stochastic dynamics of hydrogenic atoms
in the microwave field: modelling by maps and quantum
description'';

\noindent \cite{CGS87}: ``Exponential photonic localization for
the hydrogen atom in a monochromatic field'';

\noindent \cite{CGS88}: ``Hydrogen atom in monochromatic field:
chaos and dynamical photonic localization''.

\medskip

In the astronomical and physical papers by \cite{P86},
\cite{CV86}, \cite{SZ87}, \cite{PB88}, \cite{CV89}, \cite{GK87},
\cite{CGS87}, \cite{CGS88}, \cite{JLR88}, the Kepler map was
derived almost simultaneously in astronomy and physics, by means
of calculating the increments of energy and phase. However, it
should be noted that the map derivation in the problem of the
hydrogen atom in the microwave field is much simpler than in the
restricted three-body problem in celestial mechanics, because the
adopted potential model is much simpler. The energy increment in
the former problem is expressed usually through the Anger
functions.

\section{Prehistory of the Kepler map}
\label{ckm}

The second equation of the Kepler map is based on Kepler's third
law, hence the title of the map. The third law was published in
1619 in the fifth book of {\it Harmonices Mundi} \citep{K1619} ---
continuation of {\it Mysterium Cosmographicum}. A short note on
this law appeared already in 1618, in {\it A short summary of
Copernican astronomy} --- {\it Epitome Astronomiae Copernicanae}
\citep{K1618}.

Could the Kepler map have been discovered much earlier than at the
end of 20th century? To derive analytical expressions for the
energy parameter of the Kepler map, \cite{P86} and \cite{PB88}
used refined methods of mathematical physics, such as
construction of new canonical variables by means of the Lie
algebraic formalism (the Hori method),
some elements of the KAM theory, a method of reduction of a
Fourier series with a small denominator to the Fourier integral in
the form of the Cauchy integral, a method of embedding the small
denominator in an analytic function through a suitable analytic
continuation, consideration of conditions for determining the
Riemann sheet of the analytic continuation, analogies with
scattering theory in quantum mechanics. These methods became
available in the 20th century, and mostly in the sixties of the
20th century. However, the Kepler map as a mathematical
construction, put aside from the way of its original derivation,
is elementary.

The Kepler map was derived as an answer to the question, what is
the long-term orbital behaviour of comets in highly eccentric
orbits subject to perturbations from planets. It cannot be said
that this question became actual only with apparition of the
Halley comet in 1986. The highly unpredictable motion of comets is
a long-standing problem in dynamical astronomy.

As shown by \cite{V07}, Andrey \cite{L77a,L77b,L78a,L78b}
introduced the modern understanding of the dynamics of small Solar
system bodies already in 1777--1778 (this understanding implies
taking into account, first of all, the effects of resonances and
encounters with planets). Later on, \cite{L44,L48,L57}, exploring
the motion of the comet Lexell, discovered essential sensitivity
of the trajectory to the initial conditions: the trajectory
changed qualitatively upon small variations of the initial data;
this was a manifestation of the phenomenon called ``dynamical
chaos'' now \citep{V07}. Thus the scientific grounds for
exploration of the new phenomenon became actual already in the
middle of the 19th century. On the other hand, a mathematical
derivation of a formula for the energy increment in the Kepler map
(the major problem in constructing this map) could have been
accomplished since 1836, when the Jacobi integral was discovered.

It is well known that a mathematically simple setting of a problem
and a simple formulation of its solution do not at all imply a
simple way of solving the problem. What is more, arriving at
simple formulae does not at all always require simple analytical
calculations. An example can be given,
when derivation of a line-sized formula required gigabytes of
computer memory consumption \citep{S08b}. However, such a simple
mathematical construction as the Kepler map, as we have seen
above, could have well been derived, because the appropriate
scientific grounds and tools had become available, some 250 years
after the formulation by Kepler of the third law of planetary
motion, in contrast to 400 years in reality. Nevertheless the
opportunity was utterly blocked by the scientific paradigm of
Laplacian determinism.

\section{Conclusions}
\label{concl}

The Kepler map was derived in 1980s by Petrosky (1986) and
Chirikov and Vecheslavov (1986) in order to describe the long-term
chaotic orbital behaviour of comets in nearly parabolic motion.
Since that time this map has become paradigmatic in a number of
applications in celestial mechanics and atomic physics.

We have shown that the Kepler map, including analytical formulae
for its parameter, can be derived by quite elementary methods.
Though discovered so recently, it could well be derived already in
the middle of the 19th century. A strict mathematical derivation
for the energy increment could have appeared since 1836, when the
Jacobi integral was discovered.

The key word in the titles of all the pioneering papers on the
Kepler map is ``chaos'', i.e., dynamical chaos. This could not
have been a subject of a scientific study earlier than in the
second half of the 20th century. When dynamical chaos had become a
central subject of studies in nonlinear dynamics, the Kepler map
was immediately derived, together with other general separatrix
maps (Fermi map, ordinary separatrix map, Markeev maps).

I am thankful to an anonymous referee for valuable remarks
and comments.
This work was partially supported by the Russian Foundation for
Basic Research (project \# 10-02-00383)
and by the Programme of Fundamental Research of the Russian
Academy of Sciences ``Fundamental Problems in Nonlinear
Dynamics''.

\end{document}